\documentclass[%
 pre,
 reprint,
 amsmath,amssymb,
 aps,
 showpacs,
 floatfix,
 superscriptaddress,
 showkeys
]{revtex4-1}

\usepackage{graphicx}
\usepackage{dcolumn}
\usepackage{bm}
\usepackage[utf8]{inputenc}
\usepackage{hyperref}
\usepackage{paralist}

\newcommand{\ve}{\ensuremath{\varepsilon} }
\usepackage{color}

\begin{document}

\preprint{ISE15b, dated \today}

\title{Controlling Chimera States - The influence of excitable units}
%
\author{Thomas Isele}
\affiliation{Institut f\"ur Theoretische Physik, Technische Universit\"at Berlin, Hardenbergstra{\ss}e 36, 10623 Berlin, Germany}%
\email{tommaso@itp.tu-berlin.de}
\author{Johanne Hizanidis}
\affiliation{Institute of Nanoscience and Nanotechnology, National Center for Scientific Research ``Demokritos'', 15310 Athens, Greece}
\affiliation{Crete Center for Quantum Complexity and Nanotechnology, Department of Physics, University of Crete, 71003 Heraklion, Greece}
\author{Astero Provata}
\affiliation{Institute of Nanoscience and Nanotechnology, National Center for Scientific Research ``Demokritos'', 15310 Athens, Greece}
\author{Philipp H\"ovel}%
\affiliation{Institut f\"ur Theoretische Physik, Technische Universit\"at Berlin, Hardenbergstra{\ss}e 36, 10623 Berlin, Germany}%
\affiliation{Bernstein Center for Computational Neuroscience Berlin, Humboldt Universit\"at zu Berlin, Philippstr. 13, 10115 Berlin, Germany}

\date{\today}

\begin{abstract}
We explore the influence of a block of excitable units on the existence and behavior of chimera states in a nonlocally coupled ring-network of FitzHugh-Nagumo elements.
The FitzHugh-Nagumo system, a paradigmatic model in many fields from neuroscience to chemical pattern formation and nonlinear electronics, exhibits oscillatory or excitable behavior depending on the values of its parameters.
Until now, chimera states have been studied in networks of coupled oscillatory FitzHugh-Nagumo elements.
In the present work, we find that introducing a block of excitable units into the network may lead to several interesting effects.
It allows for controlling the position of a chimera state as well as for generating a chimera state directly from the synchronous state.
\end{abstract}

\maketitle


\section{Introduction}\label{sec:introduction}

Ensembles of nonlocally coupled oscillators exhibit suprising spatio-temporal patterns, called \textit{chimera states}, that consist of coexisting domains of spatially coherent (synchronized) and incoherent (desynchronized) dynamics. 
First observed in systems of identical phase oscillators with symmetric coupling topology \cite{KUR02a,ABR04}, chimera states have been intensively studied during the last decade. 
In fact, they have been observed for a wide range of local dynamics and various coupling topologies. The former include different neuronal models \cite{OLM11,OME13,HIZ13,VUE14a,OME15,HIZ15}, chaotic oscillators \cite{OME11,OME12}, Van-der-Pol systems \cite{OME15a}, Hopf normal-forms \cite{SCH14a,ZAK14,SCH15a,HAU15}, coupled rotators \cite{OLM15} to name only a few. 
Concerning the coupling, these peculiar states have been realized for different nonlocal kernels of exponential \cite{KUR02a}, sinusoidal \cite{ABR04}, hierarchical \cite{HIZ15,OME15} or rectangular shape \cite{OME10a} on a one-dimensional ring, but two-dimensional chimera states have also been reported \cite{OME12a,PAN13,PAN15a}.

Next to numerical simulations and theoretical investigations \cite{WOL11a,OME13a}, experimental evidence of chimera states was first reported in optical coupled-map lattices realized by liquid-crystal spatial light modulators \cite{HAG12} and in populations of coupled chemical oscillators \cite{TIN12}. 
In addition, these states have also been found in other settings such as mechanical experiments of two subpopulations consisting of identical metronomes \cite{MAR13}, for electrochemical oscillators \cite{WIC13}, electronic nonlinear delay oscillators \cite{LAR13}, and for superconducting meta-materials \cite{LAZ15}.
Recent research efforts aim to stabilize chimera states by feedback schemes \cite{SIE14c} and to control the localization of the different regimes, for instance, by introducing assymmetries in the coupling that drag the coherent region to one direction \cite{BIC15}.

In this study, we propose a new protocol to control the position of the coherent and incoherent regions of chimera states. It is based on modifying the system parameters of a few elements. 
In fact, we will show that a single element suffices. 
In contrast to alternative approaches that tamper with the coupling term, we leave the underlying network structure and coupling parameters untouched. 
As system of choice, we consider the paradigmatic model of the FitzHugh-Nagumo system \cite{FIT61,NAG62}, which is widely used in studying the dynamics of excitability. 
In our configuration, the nodes operate in the oscillatory regime except for the control elements, which are excitable.

The rest of the work is organized as follows: We introduce the model in Sec.~\ref{sec:model}. In Sec.~\ref{sec:effect-excit-elem}, we investigate the effects of excitable elements in an otherwise oscillatory ring with respect to the position of the chimera state. 
We will finally conclude in Sec.~\ref{sec:conclusion}.

\section{Model}\label{sec:model}
We consider a one-dimensional ring of $N$ nonlocally coupled FitzHugh-Nagumo units where every node is coupled to its $R$ neighbors on either side \cite{OME11,OME12,OME13,OME15}.
\begin{subequations}\label{eq:fhn_on_ring}
\begin{align}
  \ve\dot u_i    =&   u_i - \frac{u_i^3}{3} - v_i\nonumber\\
  &+ \frac{\sigma}{2R} \sum_{|j-i|\le R}\left[c_{uu}(u_j-u_i) + c_{uv}(v_j-v_i)\right]\label{eq:fhn_on_ring_u}\\
  \dot v_i       =&   u_i + a_i                  + \frac{\sigma}{2R} \sum_{|j-i|\le R}\left[c_{vu}(u_j-u_i) + c_{vv}(v_j-v_i)\right],\label{eq:fhn_on_ring_v}
\end{align}
\end{subequations}
where $u_i$ and $v_i$, $i=1,\dots,N,$ are the activator and inhibitor variables, respectively, and indices are to be understood modulo $N$.
The parameter \ve separates the timescales of the fast activator and slow inhibitor dynamics.
Throughout this work, we choose a fixed value of $\ve=0.05$.
Depending on the value of the threshold parameter $a_i$, an individual unit shows either oscillatory ($|a_i|<1$) or excitable ($|a_i|>1$) behavior.
The parameter $\sigma$ denotes the coupling strength.
Another important feature of the coupling of Eqs.~\eqref{eq:fhn_on_ring} is the presence of cross-coupling terms between the activator and inhibitor variables.
This is modeled by a rotational matrix
\begin{align}
{\bf B} &= \begin{pmatrix} c_{uu}&c_{uv}\\c_{vu}&c_{vv}\end{pmatrix} =
\begin{pmatrix} \cos\phi&\sin\phi\\-\sin\phi&\cos\phi\end{pmatrix},\label{eq:coupling_matrix}
\end{align}
which is determined by a single coupling phase $\phi$.
In Ref.~\cite{OME13}, the existence of chimeras in Eqs.~\eqref{eq:fhn_on_ring} for a homogeneous system of nodes in the oscillatory regime was shown for coupling phases slightly below $\pi/2$.
In this work we choose the same value $\phi=\pi/2 -0.1$.

To quantify the existence and properties of a chimera state, two widely used indicators are employed.
The mean phase velocity is defined as 
\begin{align}
  \omega_i    &= \left\langle \frac{\text{d}\theta_i(t)}{\text{d}t}\right\rangle_{\Delta t},\label{eq:average_phase_velocity}
  \intertext{where  $\theta$ is the geometric angle in the $(u_i,v_i)$-plane}
  \theta_i(t) &= \arctan\left(\frac{v_i(t)}{u_i(t)}\right),  
\end{align}
and $\Delta t$ is the time window over which the average is computed.
The mean phase velocity profile of a chimera state is given by a curve, which is flat in the coherent region of the chimera and arc-shaped in the incoherent one.
To obtain a smooth mean phase velocity profile it is necessary to average over a large time $\Delta t$ spanning many thousand periods.
This makes it difficult to monitor fast changes in the coherent and incoherent regions such as drift of the chimera state.

\begin{figure}[tb]
\includegraphics{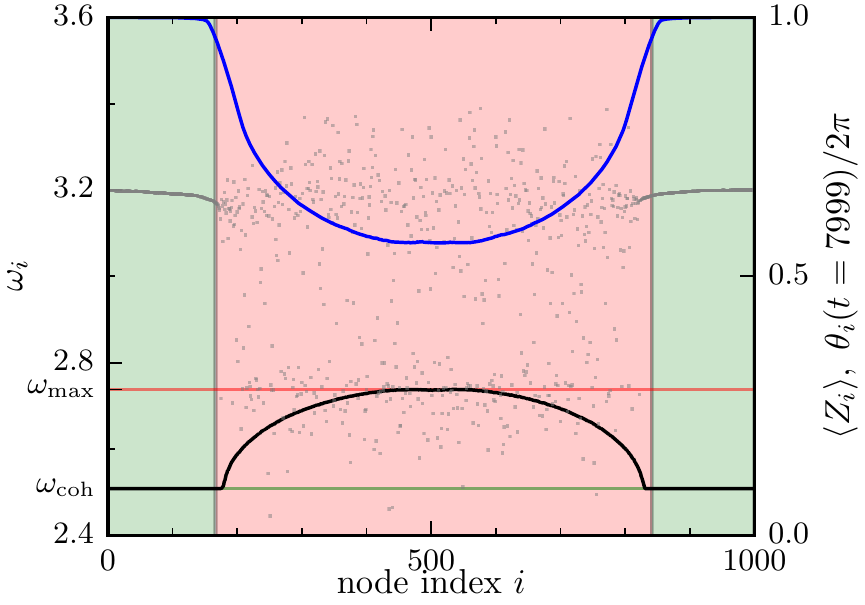}
\caption{\label{fig:chimera_didactical}
(Color online) A typical chimera state in system Eqs.~\eqref{eq:fhn_on_ring} under homogeneous conditions $a_i\equiv 0.5$ (i.e. $b=0$).
Shown are the mean phase velocity $\omega_i$ (black)
the (temporally) averaged local order parameter $\langle Z_i \rangle$ (blue) and a snapshot of the angles $\theta_i$ (gray).
The incoherent region is marked by a red background, the coherent region is marked by a green background.
Horizontal lines mark 
the frequency of the homogeneous region $\omega_\text{coh}$ (green) and 
the maximum frequency in the incoherent region $\omega_\text{max}$ (red).
The time window for calculating $\omega_i$ and $\langle Z_i \rangle$ is $\Delta t=1000$ (from the end of the timeseries). 
The spatial window for calculating $Z_i$ is $\delta=25$.
Other parameters are $N\!=\!1000,\ R\!=\!350,\ \sigma\!=\!0.15,\ \phi\!=\!\pi/2-0.1,\ \ve\!=\!0.05$.}
\end{figure}

A measure that can be calculated at every instant of time indicating the spatial coherence of a region, is the so-called \emph{local order parameter}.
It is defined as
\begin{align}
  Z_i(t) &= \left|\frac{1}{2\delta}\sum_{|j-i|\le\delta}e^{\mathrm{i} \Theta_j(t)}\right| , \label{eq:local_order_parameter}
\end{align}
where $\delta$ is the spatial window size. 
We use $\delta\!=\!25$ throughout this work.
To account for slightly non-synchronous transitions between the left and right branches of the cubic nullcline, which arise from the slow/fast character of the dynamics, we use the moving average of the local order parameter which is calculated over a (small) time window $\Delta t$:
\begin{align}
\left\langle Z_i\right\rangle_{\Delta t}(t)  &= \frac{1}{\Delta t} \int_{t-\Delta t}^{t} Z_i(s) \mathrm{d}s. \label{eq:moving_average_of_Z}
\end{align}

Fig.~\ref{fig:chimera_didactical} shows a typical chimera state by means of these two indicators as well as a snapshot of the activator values $u_i$.

\begin{figure}[htbp]
\includegraphics{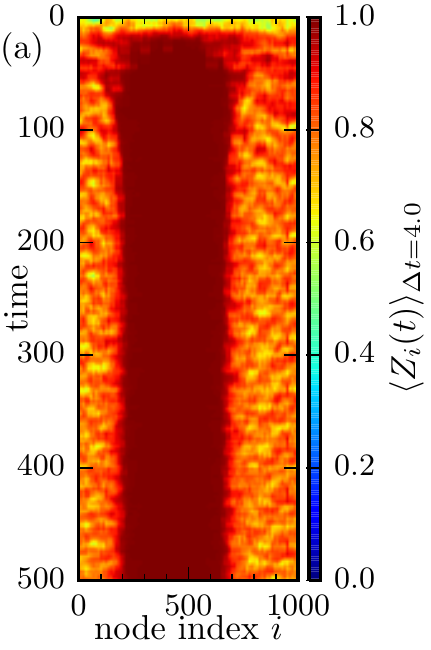}%
\includegraphics{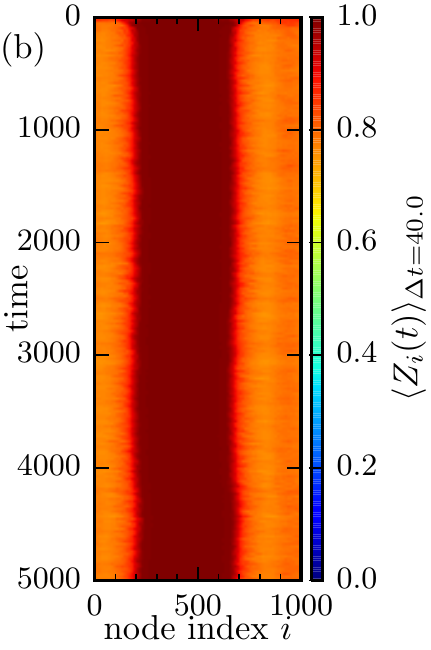}
\caption{\label{fig:ts_plot_nucleation_homogeneous}(Color online) Space-time plot of the local order parameter $\langle Z_i(t)\rangle_{\Delta t}$ for (a) $t\in(0,500)$ and (b) $t\in(0,5000)$ for system Eqs.~\eqref{eq:fhn_on_ring} with barrier width $b=0$ starting from random initial conditions as described in Sec.~\ref{sec:effect-excit-elem}.
Other parameters are $N\!=\!1000,\ R\!=\!350,\ \sigma\!=\!0.2,\ \phi\!=\!\pi/2-0.1,\ a_\text{osc}\!=\!0.5,\ \ve\!=\!0.05$.}
\end{figure}

In this work, we investigate the influence of localized parameter inhomogeneities.
To this end, we choose a value in the oscillatory regime for all nodes except for one connected region of a few nodes that operate in the excitable regime.
To be precise, we select the threshold parameters as follows:
\begin{align}
  a_i &= 
\begin{cases} 
a_\text{osc}   & \text{for }i  >  b \\
a_\text{exc}   & \text{for }i \le b
\end{cases},
\end{align}
where $a_\text{osc}$ is a (fixed) value in the oscillatory regime, $a_\text{exc}$ is a (fixed) value in the excitable regime and $b$ is the number of excitable units.
We call $b<N$ the \emph{barrier width} and $a_\text{exc}$ the \emph{barrier height}, thus employing the picture of a barrier of excitable units located at indices $i=0$ to $b$.
Note that by making an index shift (rotation along the one-dimensional ring), the barrier can be placed at an arbitrary location.

\section{Effect of excitable units}\label{sec:effect-excit-elem}
It has been demonstrated in Ref.~\cite{OME13} that for appropriate choice of the coupling matrix~\eqref{eq:coupling_matrix} and the parameters $R$ and $\sigma$, chimera states can emerge in system Eqs.~\eqref{eq:fhn_on_ring} with homogeneous threshold parameter $a_i\equiv a$ in the oscillatory regime.
In such a setup, the location of the incoherent region is determined by the initial conditions alone, as the system itself is invariant against rotations (index shifts).
Initial conditions typically employed in this setting are uniformly randomly distributed values of $(u_i,v_i)$ on a circle of radius 2, that is, $u_i^2+v_i^2=4$, or completely random.

\begin{figure}[tb]
\includegraphics{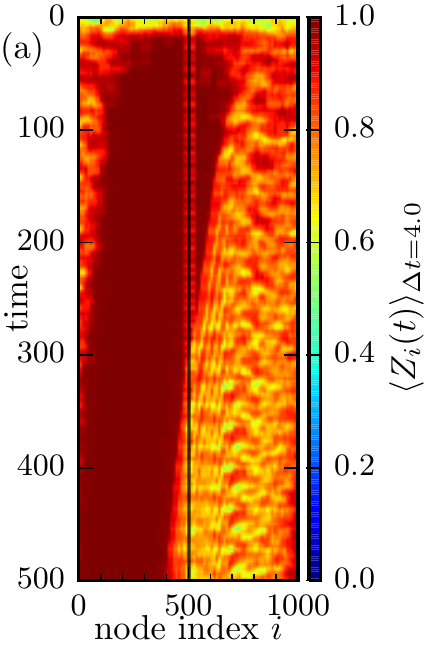}%
\includegraphics{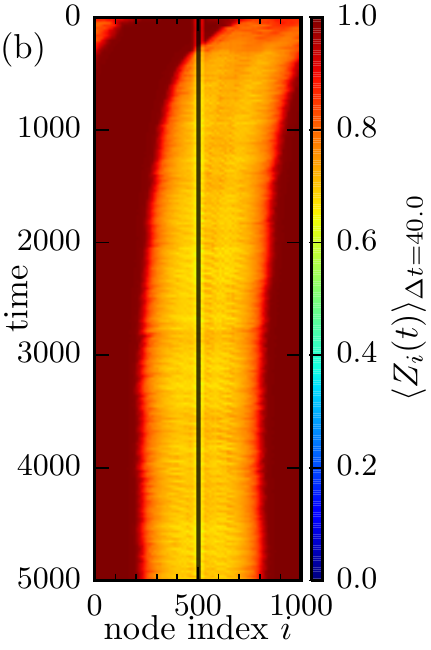}
\caption{\label{fig:ts_plot_nucleation_barrier}(Color online) Space-time plot of the local order parameter $\langle Z_i(t)\rangle_{\Delta t}$ for (a) $t\in(0,500)$ and (b) $t\in(0,5000)$ for system Eqs.~\eqref{eq:fhn_on_ring} with barrier width $b=5,\ a_\text{exc}=1.3$ starting from the exact same initial conditions as in Fig.~\ref{fig:ts_plot_nucleation_homogeneous}.
The location of the barrier of excitable elements is marked by vertical black lines.
Other parameters are as in Fig.~\ref{fig:ts_plot_nucleation_homogeneous}}
\end{figure}

\subsection{Attraction of incoherent region to barrier}\label{sec:attr-incoh-reg}

Figure~\ref{fig:ts_plot_nucleation_homogeneous} depicts a space-time plot of $\langle Z_i(t)\rangle$ for Eqs.~\eqref{eq:fhn_on_ring} without a barrier of excitable elements, $a_i \!\equiv\! a_\text{osc}\!=\!0.5$.
Coherent and incoherent regions clearly form after a short time ($t\approx 50$) with positions determined solely by the initial conditions (see Fig.~\ref{fig:ts_plot_nucleation_homogeneous}(a)).
Over the course of the simulation, the sizes, locations and average values of the local order parameter for the coherent and incoherent region hardly change (see Fig.~\ref{fig:ts_plot_nucleation_homogeneous}(b)). Note that up to $t=5000$ the timeseries comprises approximately 2000 periods.

The situation is somewhat different, when a barrier of excitable elements is introduced.
In Fig.~\ref{fig:ts_plot_nucleation_barrier}, a similar density plot is shown except that here we consider a barrier width of $b=5$ and $a_\text{exc}=1.3$, which has been shifted to node 500 for visualization purposes.
The initial conditions used in this simulation are the exact same as those employed for Fig.~\ref{fig:ts_plot_nucleation_homogeneous}.
Indeed, a close look at Fig.~\ref{fig:ts_plot_nucleation_barrier}(a) and comparison with Fig.~\ref{fig:ts_plot_nucleation_homogeneous}(a) reveals that, in the beginning when the chimera starts to form (for about 100 time units), both space-time plots resemble each other almost perfectly.

After this short interval, the presence of the excitable units shows a suprising effect:
The size of the coherent and incoherent region does not change significantly but the regions begin to drift in such a way that in the end, the barrier of excitable elements finds itself in the middle of the incoherent region (rather than in the coherent region, where it was originally).
The barrier of excitable units effectively attracts the incoherent region (see Fig.~\ref{fig:ts_plot_nucleation_barrier}(b)).
Furthermore, the average value of the local order parameter in the entire incoherent region is lower in the presence of the barrier.
This effect is only partially caused by the finite size of the spatial window ($\delta=25$) used to calculate the local order parameter (Eq.~\eqref{eq:local_order_parameter}) as this would only influence nodes at a maximum distance $\delta$ from the barrier.

\begin{figure}[tb]
\includegraphics[width=0.5\columnwidth]{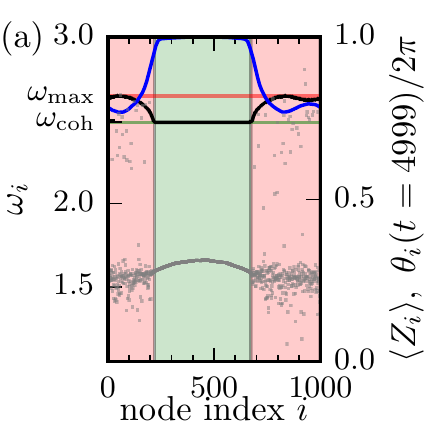}%
\includegraphics[width=0.5\columnwidth]{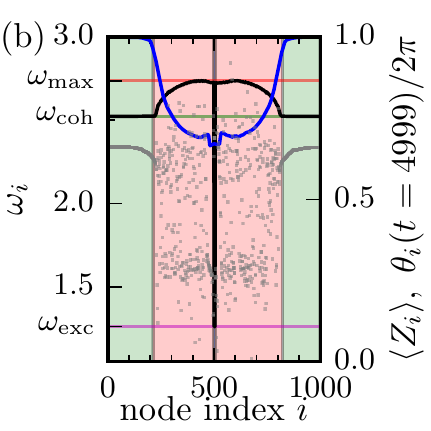}
\caption{\label{fig:omega_av_plot}(Color online) 
Mean phase velocity $\omega_i$ (black, left scale),
averaged local order parameter $\langle Z_i \rangle$ (blue) and a snapshot of the phases $\theta_i$ (grey) for the chimera states obtained for
(a) $b=0$ and (b) $b=5,\ a_\text{exc}=1.3$. 
The incoherent region is marked by a red background, the coherent region is marked by a green background.
Horizontal lines mark 
the frequency of the coherent region (green), 
maximum frequency in the incoherent region (red) and 
average frequency of the excitable units (magenta, only(b)).
The time window for calculating $\omega_i$ and $\langle Z_i \rangle$ is $\Delta t=1000$ (from the end of the timeseries). 
Other parameters are as in Fig.~\ref{fig:ts_plot_nucleation_homogeneous}.}
\end{figure}

We also observe a slight increase of the mean phase velocity in the coherent as well as in the incoherent regions.
This is shown by $\omega_\text{coh}$ and $\omega_\text{max}$ in Fig.~\ref{fig:omega_av_plot}(a) and~(b), where $\omega_i$ is plotted for the last 1000 time units of the simulations of Fig.~\ref{fig:ts_plot_nucleation_homogeneous}~and~\ref{fig:ts_plot_nucleation_barrier}, respectively.

\subsection{Controlling the location of the chimera}\label{sec:contr-locat-chim}
In the previous section, we showed that even a very small number of excitable units is sufficient to attract the incoherent region.
In order to reach an effective control of the location of the position of the chimera, we assess the following questions
\begin{inparaenum}[(i)]
  \item How long does it take until the final position is reached by the incoherent region?
  \item How does the final position depend on  the position of the barrier of excitable elements?
  \item How is the size of the chimera (defined as the size of the incoherent region) influenced? 
\end{inparaenum}

\begin{figure}[tb]
\includegraphics{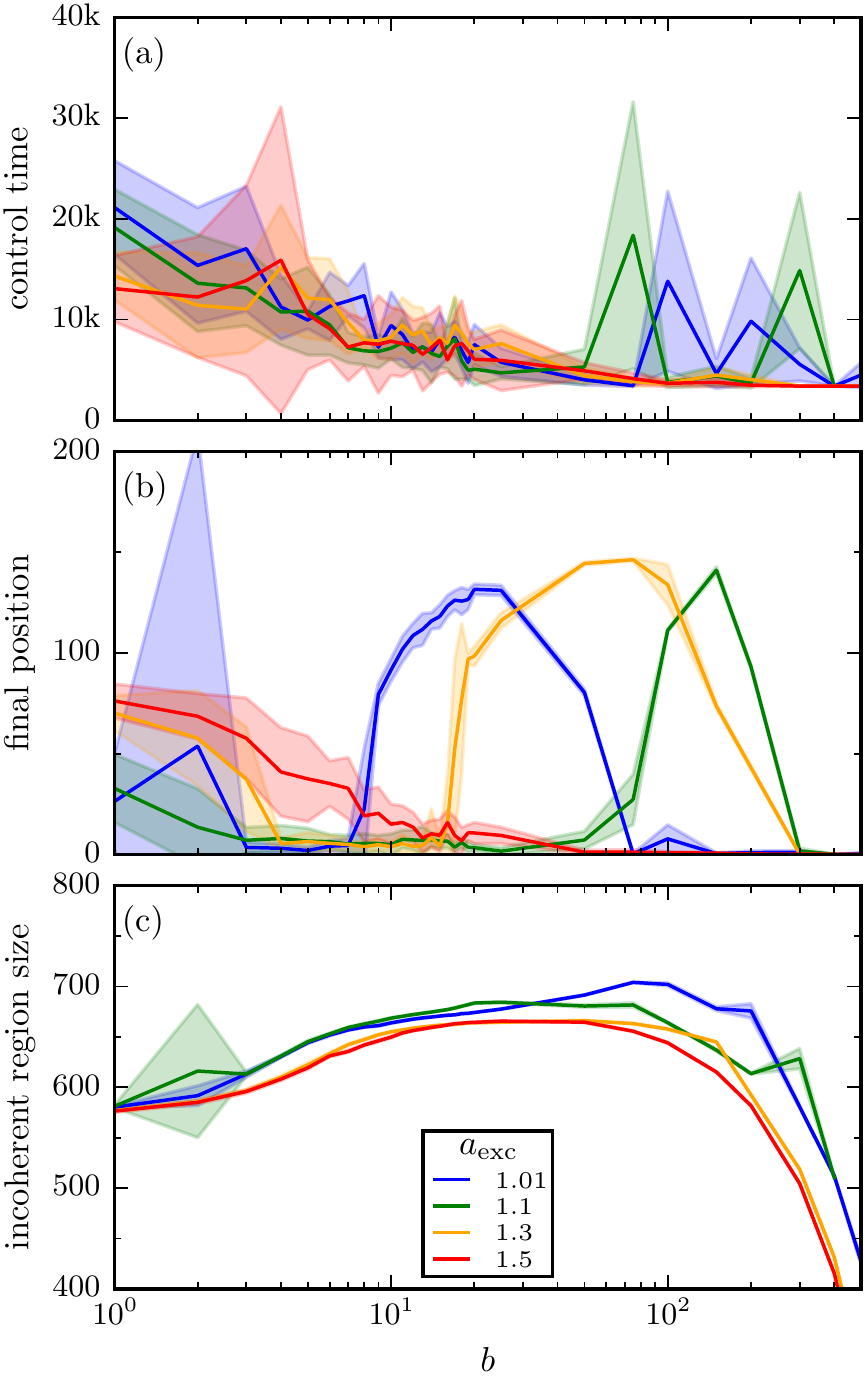}%
\caption{\label{fig:control_efficiency}
(Color online) Measures of control efficiency vs. barrier size $b$ for  barrier heights $a_\text{exc}=1.01,\,1.1,\,1.3,$ and $1.5\,$.
Shown is the average value (lines in lucid colors) and the standard deviation (transparent areas below the lines).
(a) control time, 
(b) final position.
(c) size of incoherent region
The statistics are taken for chimeras generated for $N=1000,\ R=350,\ \sigma=0.2$ from 10 different initial conditions as described in Sec.~\ref{sec:attr-incoh-reg}.
Other parameters as in Fig.~\ref{fig:ts_plot_nucleation_homogeneous}. }
\end{figure}

For this purpose, we generate chimera states from different initial conditions for fixed values of $R=350$ and $\sigma=0.2$. 
Using a snapshot of these chimera states as initial conditions, we perform a numerical simulation with a barrier of width $b$ and height $a_\text{exc}$ of excitable elements, which is placed exactly opposite from the center of the incoherent region.
During the simulation, we monitor the boundaries between incoherent and coherent region using the moving average of the local order parameter (Eq.~\eqref{eq:moving_average_of_Z}).
If the speed with which the center of the incoherent region is moving falls under a certain threshold (taken to be 1 node per 2000 time units), and the rate of change of size of the incoherent region becomes less than 1 node per 1000 time units, the simulation is stopped.
The time at which both threshold values are achieved is the \emph{control time}.
The distance (measured in nodes) between the center of the incoherent region and the center of the barrier gives the \emph{final position}.
The final size of the incoherent region without counting the excitable elements is also measured.
The values for these three quantities averaged over 10 different chimeras for $R=350$ and $\sigma=0.2$ is shown in Fig.~\ref{fig:control_efficiency}, also including the standard deviation (displayed as transparent areas).
The width of the barrier has been varied between $b=1$ and $b=500$. 
For values $b>500$ we could not find a pronounced enough difference between coherent and incoherent region in most of the cases.
The procedure of measuring the drift velocity is described in detail in Appendix~\ref{sec:prot-meas-chim}.

In Fig.~\ref{fig:control_efficiency}, we plot the three measures of coherence/incoherence that evaluate the effect of the control vs. the width $b$ of the barrier.
We do this for four barrier heights $a_\text{exc}=1.01,\,1.1,\,1.3,$ and $1.5$.
Fig.~\ref{fig:control_efficiency}(a) shows the average value of the control time, (b) shows the final position and  (c) shows the incoherent region size.
The curves of these values are surrounded by a pale area, which indicates the size of the standard deviation.
The behavior of the control time, as one would naively expect, decreases with increasing barrier width $b$ as well as with increasing barrier height $a_\text{exc}$.
So does the final position at low values of the barrier width (up to approximately $b=5$).
An exception is the curve for $a_\text{exc}=1.01$ which, however, also exhibits a very large standard deviation.

For increasing barrier width, the final position of the chimera lines up increasingly well with the position of the barrier.
For larger barriers the final position shows a peculiar behavior.
For barrier heights below $a_\text{exc}=1.4$, there is a range of barrier widths $b$ in which the incoherent region does not line up exactly with the position of the barrier.
For these values, the final position shows a maximum at a certain value of $b$. 
The barrier width at which this maximum is attained does not show a simple dependence on the value of $a_\text{exc}$.
The small standard deviation at the location of this phenomenon shows that it can be consistently reproduced in our setup.
The size of the incoherent region becomes somewhat larger with increasing barrier width until it drops to zero at very large barrier widths.

\begin{figure}[tb]
\includegraphics[width=0.5\columnwidth]{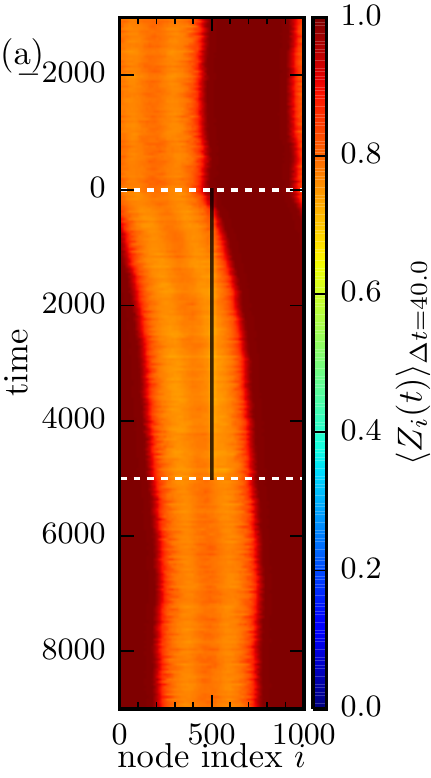}%
\includegraphics[width=0.5\columnwidth]{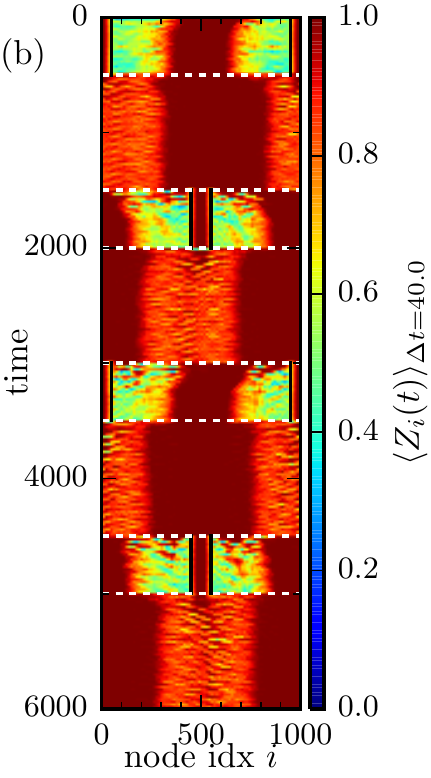}
\caption{\label{fig:on_off_and_chessboard}
(Color online) Space-time plot of $\langle Z_i(t)\rangle_{\Delta t}$.
(a) $b=1,\ a_\text{exc}=1.1$ from $t=0$ until $t=5000$.
Before and after that $b=0$.
(b) Steering the chimera with $b=100,\ a_\text{exc}=1.5$. 
For details on the protocol see Sec.~\ref{sec:contr-locat-chim}.
Switching times are marked by white dashed lines.
Location of excitable units is marked by black solid lines.
Other parameters as in Fig.~\ref{fig:ts_plot_nucleation_homogeneous}}
\end{figure}

Once the steady chimera pattern is realized in the presence of the barrier, after resetting the excitable units back to oscillatory and thus rendering the system homogeneous again, the incoherent region remains stationary.
This is demonstrated in Fig.~\ref{fig:on_off_and_chessboard}(a), where we first generate a chimera state in a homogeneous system of size $N\!=\!1000$, then at some point in time (labeled $t\!=\!0$) we change the threshold parameter of a single node to be excitable.
After 5000 time units, we change the threshold parameter again to the old value in the oscillatory regime.
Within this 5000 time units, we observe that the incoherent region drifts in the way described in Sec.~\ref{sec:attr-incoh-reg}.
After switching off the inhomogeneities, i.e., returning to the homogenous system, the location of the incoherent region remains the same.

In the example of Fig.~\ref{fig:on_off_and_chessboard}(a), the drift of the chimera is slow compared to the intrinsic timescale.
When size $b$ and height $a_\text{exc}$  of the barrier grow larger, the drift happens faster.
Using a barrier size $b=100$ and barrier height $a_\text{exc}=1.5$, for instance, we demonstrate steering of the chimera in Fig.~\ref{fig:on_off_and_chessboard}(b).
In a homogeneous system with a chimera solution, we switch on the barrier for 500 time units in one location, then off for 1000 time units, then on again in a different location for another 500 time units and off again for 1000 time units.
The two locations are placed on opposite sides of the ring network and the procedure is repeated.
In Fig.~\ref{fig:on_off_and_chessboard}(b), the local order parameter shows that this protocol can be used to efficiently control the position of the chimera in the system.
Moreover, the figure shows that at such a large width and height of a barrier, the incoherent region grows significantly and the value of the local order parameter in the incoherent region is also reduced considerably.
Both effects vanish as soon as the barrier is switched off.

For the chosen parameter values, the system in the absence of a barrier exhibits a 1-chimera state.
Control by the barrier can also used for multichimera states. 
In those cases, typically one of the incoherent regions of the multichimera state is attracted to the barrier while the other incoherent regions drift along.
In the final state, this one region sits on top of the barrier and the others are next to it.

\subsection{Generation of incoherent regions due to excitable units}\label{sec:gener-incoh-regi}
\begin{figure}
\includegraphics[width=\columnwidth]{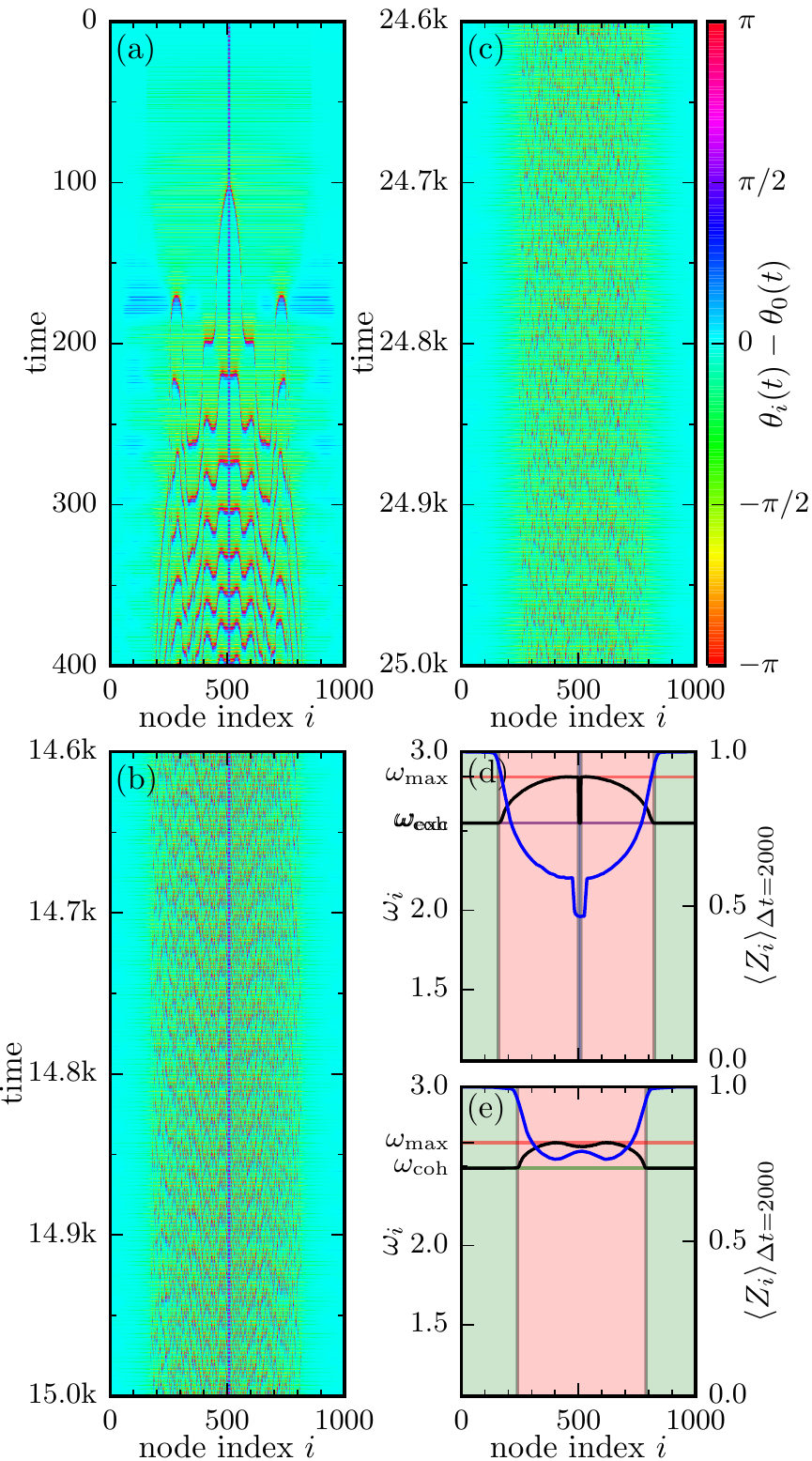}
\caption{\label{fig:gen-incoh-reg}
(Color online)
Space-time plot of $\theta_i-\theta_{0}$, starting with homogeneous initial conditions at $t=0$. 
Between $t=0$ and $t=15000$, $b=10,\,a_\text{exc}=1.1$. From $t=15000$, $b=0$.
(a) $t=0-400$,
(b) $t=14600-15000$,
(c) $t=24600-25000$.
(d) $\omega_i(t=15000)$ and $\langle Z_i \rangle(t=15000)$ with $\Delta t=2000$ for the calculation of both.
(e) Same as (d) for $t=25000$.
The excitable units are located at $i\in(500,509)$.
The timeseries is showing sudden phase differences drifting into the oscillatory region, leading to a chimera-like state after a large time, which persists after the barrier is turned off again.
Other parameters are as in Fig.~\ref{fig:ts_plot_nucleation_homogeneous}.}
\end{figure}

In Eqs.~\eqref{eq:fhn_on_ring} with a homogeneous parameter setup i.e. $b=0$, the synchronous solution is always stable \cite{OME13}.
When integrating the system with homogeneous initial conditions, the system will always end up in this stable synchronized state.
However in a setup including a barrier of excitable elements, we have found a different result.
First, in the system including the barrier, there need not be a synchronous state because of the different regimes (oscillatory and excitatory) of the elements.
To demonstrate the behavior of the system including a barrier of excitable units, we start with a synchronous oscillation in the homogeneous system.
At one instant of time (taken to be $t=0$) we turn on the barrier of excitable elements.
Originating at the border between excitable and oscillatory units, sudden phase differences between neighboring oscillatory units emerge and drift into the oscillatory region (see Fig.~\ref{fig:gen-incoh-reg}(a)).
These phase differences do not travel through the entire system but accumulate at certain distances left and right of the barrier.
After some time, a large number of these phase differences have accumulated in an entire region around the excitable elements.
Even though the phase differences start out symmetrically from the barrier, the state after a long time is not symmetrical (around the barrier) anymore (see Fig.~\ref{fig:gen-incoh-reg}(b)).
This might be due to computational noise as inevitably induced by rounding errors in the simulation and needs further investigation.

The region with accumulated phase differences shows all the signs of a typical incoherent region in a chimera state.
The mean phase velocity $\omega_i$ is elevated and the local order parameter decreases in this region.
Even after setting the threshold parameter of the barrier back to the oscillatory value, i.e. turning off the barrier, this incoherent region persists and cannot be distinguished from a normal chimera state.

An exemplary timeseries is shown in Fig.~\ref{fig:gen-incoh-reg}.
Here, we choose $R=350$ and $\sigma=0.2$.
The barrier is turned on at $t=0$ and we simulate the system with a barrier until $t=15000$. 
After that we turn off the barrier and simulate the system for another 10000 time units until $t=25000$.
In Fig.~\ref{fig:gen-incoh-reg}(a)-(c), different stages of the simulation are shown as a space-time plot.
We shifted the indices $i$ so the barrier is between nodes 500 and 509. 
Displayed in these figures is the difference $\theta_i-\theta_0$, where $\theta_0$ is the phase of the oscillator in the middle of the coherent region defining thus a co-rotating system.
In Figs.~\ref{fig:gen-incoh-reg}(d) and~(e), $\omega_i$ and $\langle Z_i \rangle$ are shown averaged over the last 2000 time units of (d) the system with barrier switched on and (e) the system with barrier switched off.
One clearly sees that the state in which the system remains shows exactly the same characteristics as the uncontrolled chimera state (cf. Fig.~\ref{fig:omega_av_plot}(a)).
This can be explained as follows: The introduction of the barrier creates an inhomogeneous state and hence breaks the symmetry.
If this state is chosen as initial condition for the uncontrolled system, the formation of chimera states is favoured and therefore a chimera state emerges.

\subsection{Effect of excitable units in the parameter plane}\label{sec:effect-excit-units}
To provide further insight into the effects of a barrier of excitable elements on the occurrence and properties of chimera states, we vary the coupling range $R$ and the coupling strength $\sigma$.
We start with a chimera solution at $R=350$ and $\sigma=0.2$, track this solution by slightly changing the parameters $R$ and $\sigma$ and letting the numerical simulation run with (the last snapshot of) the previous chimera state as initial conditions.
This way we scan the parameter plane $(R,\sigma)$. 
We consider four different configurations of the barrier: $b=10,\,100$ with both $a_\text{exc}=1.1$ and $a_\text{exc}=1.5$.

\begin{figure}[tb]
\includegraphics[width=\columnwidth]{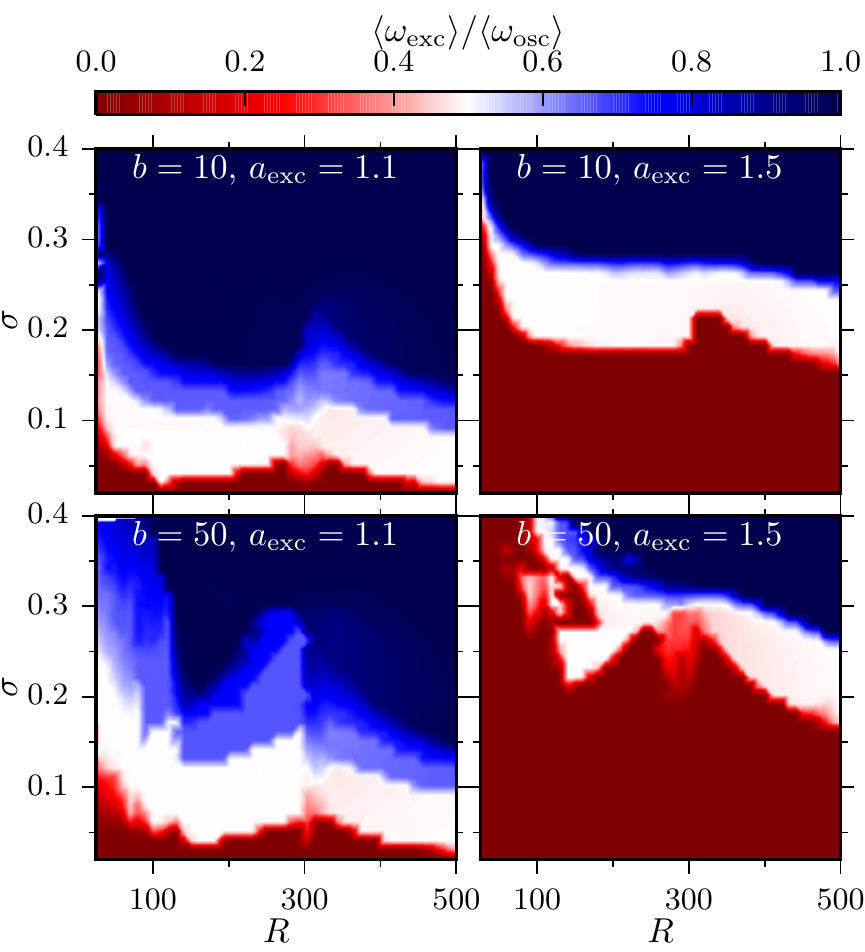}
\caption{\label{fig:omega_ratio}(Color online) 
Space-time plot of the ratio of the average frequencies of excitable and oscillatory nodes $\langle \omega_\text{exc} \rangle / \langle \omega_\text{osc} \rangle$ as a function of the parameters $R$ and $\sigma$.
Values of $b$ and $a_\text{exc}$ are given on the panels.
Other parameters are as in Fig.~\ref{fig:ts_plot_nucleation_homogeneous}}
\end{figure}

\begin{figure}[tb]
\includegraphics[width=\columnwidth]{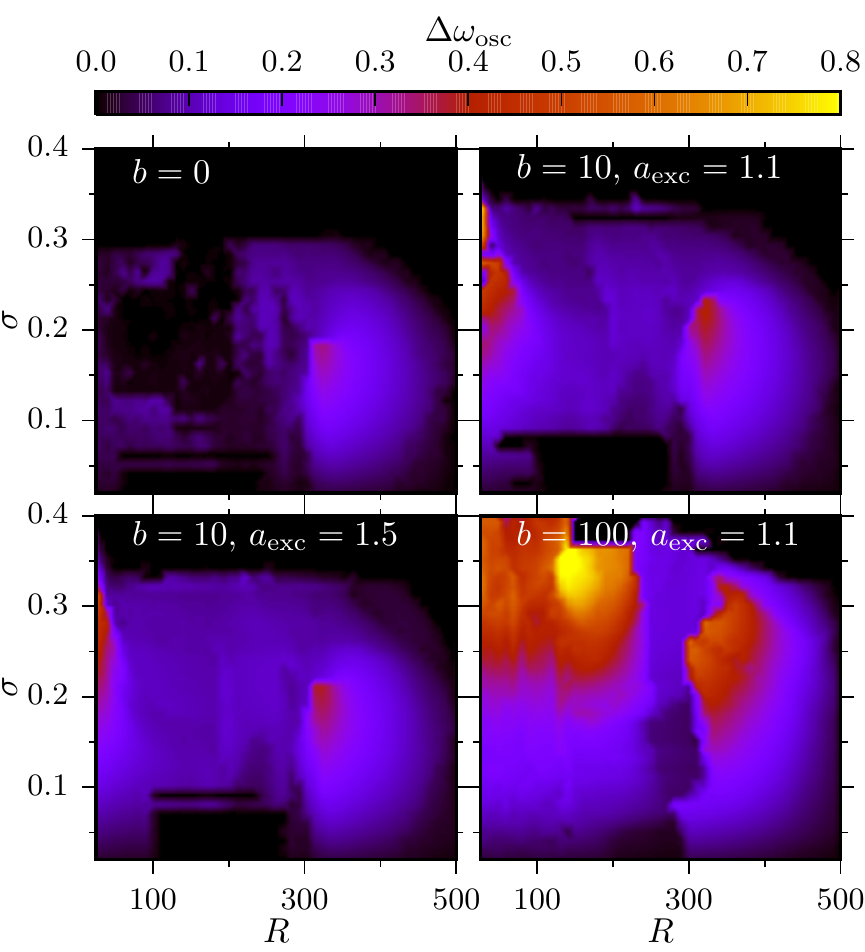}
\caption{\label{fig:Delta_omega}(Color online) 
Space-time plot of $\Delta \omega_\text{osc}$ as a function of the parameters $R$ and $\sigma$. 
Values of $b$ and $a_\text{exc}$ are given on the panels.
Other parameters are as in Fig.~\ref{fig:ts_plot_nucleation_homogeneous}}
\end{figure}

\begin{figure}[tb]
\includegraphics[width=\columnwidth]{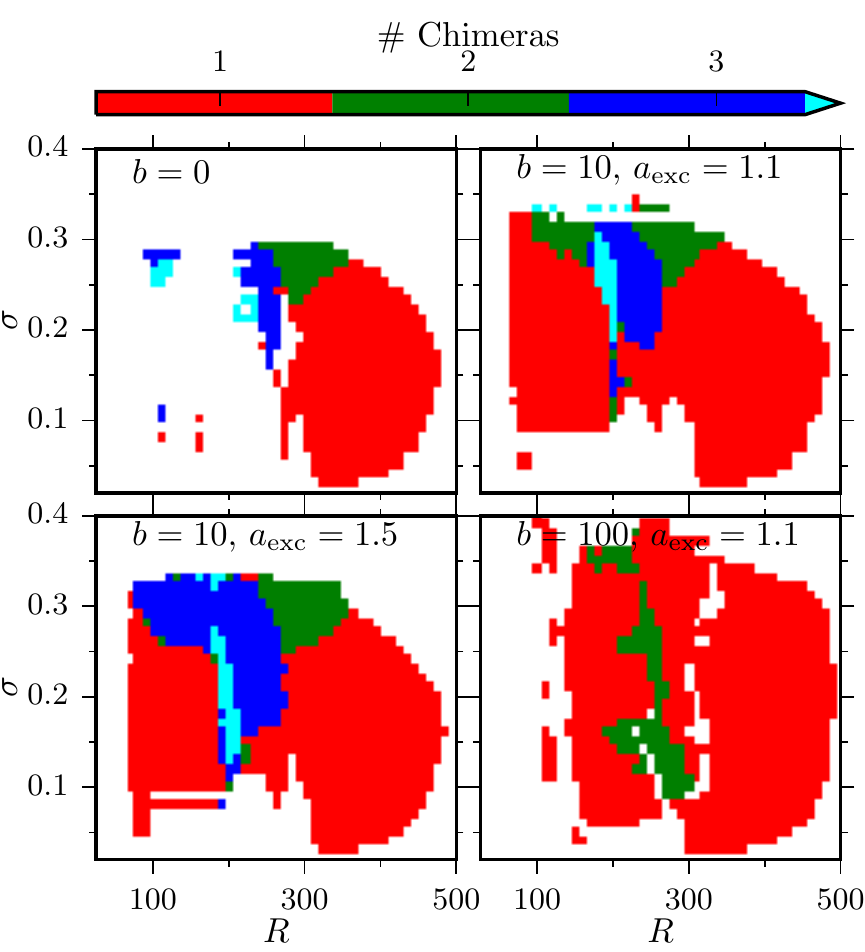}
\caption{\label{fig:num_chim}(Color online) 
Number of incoherent regions (chimera index) calculated by the algorithm described in the text.
Values of $b$ and $a_\text{exc}$ are given on the panels.
Other parameters are as in Fig.~\ref{fig:ts_plot_nucleation_homogeneous}}
\end{figure}

There are several quantities that give clues about the behavior of chimera states in our system.
Figure~\ref{fig:omega_ratio} shows the ratio   $\langle \omega_\text{exc} \rangle / \langle \omega_\text{osc}\rangle$, where $\langle \omega_\text{exc}\rangle$ and $\langle \omega_\text{osc}\rangle$ are the mean phase velocities in the excitable and oscillatory regime, respectively.
In general, increasing the coupling strength increases the mutual coupling between excitatory and oscillatory nodes and thus leads to the excitatory nodes participating in more of the oscillations of the oscillatory units which leads to an increase of $\langle \omega_\text{exc} \rangle / \langle \omega_\text{osc} \rangle$.
This increase, however, is by no means linear.
Values close to $\langle \omega_\text{exc} \rangle / \langle \omega_\text{osc} \rangle=0,\,0.5,\,1$ are preferred by the system, as is clearly seen in Fig.~\ref{fig:omega_ratio}. $\langle \omega_\text{exc} \rangle / \langle \omega_\text{osc} \rangle=0$ refers to the subthreshold oscillations of the excitable elements.
Also, the transition points in $\sigma$ show a strong dependence on $R$ in all of the barrier configurations.
The general behavior seems to be that the transition points are shifted to lower $\sigma$ with increasing $R$. 

The difference between the lowest and the highest frequency of the oscillatory nodes is denoted by $\Delta \omega_\text{osc}$ and is shown in Fig.~\ref{fig:Delta_omega} for the four barrier configurations.
The effect of increasing the barrier width $b$ is clearly visible as an increase of $\Delta \omega_\text{osc}$ in the regime of low coupling range $R$.

We have designed an algorithm that examines a timeseries and computes 
\begin{inparaenum}[(a)]
  \item whether the system is in a chimera state and
  \item if it is, how many incoherent regions are present (chimera index).
\end{inparaenum}
This algorithm works by evaluating the local order parameter and the mean phase velocity.
The algorithm is explained in detail in Appendix~\ref{sec:algor-detect-chim}.
The reason to adopt a more involved procedure is that some marginal cases are hard to treat, even by manual inspection. 
The results of the algorithm for the four barrier configurations is shown in Fig.~\ref{fig:num_chim}.
The effect of the barrier manifests itself in different ways regarding the influence of barrier height $a_\text{exc}$ and barrier width $b$.
A larger barrier height increases the area where  multi-chimera states occur as seen by inspecting Fig.~\ref{fig:num_chim}.
Increasing the barrier width, however, significantly enlarges the area where one-headed chimera states occur not only at the expense of multi-chimera states but also recruiting areas where no chimera states have been found with smaller barriers (or without a barrier).

Especially in the region of small coupling range $R$ ($R<200$), chimera states are detected in the system with barrier (Fig.~\ref{fig:num_chim}(b)-(d)) while in the system without barrier (Fig.~\ref{fig:num_chim}(a)), no chimera states have been found.
The chimera states that we found here have the peculiarity that in the coherent region there is a constant phase lag between the oscillators and thus the coherent region does not show synchronous oscillations but (counter-)propagating waves.


\section{Conclusion}\label{sec:conclusion}
We have studied the influence of a block (or barrier) of excitable units on chimera states in a nonlocally coupled ring of otherwise oscillatory FitzHugh-Nagumo systems. 
The width of the barrier has been changed from a single element up to 50\% of the total number of nodes.
In this study, only the system parameters were modified, the coupling topology and the coupling scheme remaining untouched.

We have observed several notable effects. 
Generally speaking, the modification of the system parameters facilitates the occurrence of chimera states in the network. 
We have identified chimera states in regions of the parameter plane of coupling range and coupling strength where, without the barrier, no chimeras have been observed. 
Depending on the barrier width (number of units in the block) and height (value of the threshold parameter within the block), the occurrence of multi-chimera states can be facilitated (small width, large height) or suppressed (large width). 
At small values of the coupling range, we also found chimera states in which the coherent region shows traveling wave behavior, that is, the elements in the coherent region are frequency-locked, but exhibit a constant phase lag between each other. 
This has, to the best of our knowledge, not been reported before.

Moreover, the barrier influences the position of the incoherent region of a chimera state. 
In the presence of a barrier of excitable elements, the incoherent region drifts towards it, until the excitable units are in its interior, often perfectly centered. 
This provides an interesting tool to deliberately control the chimeras' position by small modifications of the system. 
The presented scheme is different from other control schemes like the one proposed in Ref.~\cite{BIC15} in which the coupling topology (or at least the coupling weights) of the entire network has to be altered. 
In our setup, we have found that, as little as one single excitable unit, which operates only just beyond the bifurcation point in the excitable regime, suffices to steer the incoherent region of a chimera state.

Another interesting observation is that such a barrier not only facilitates the existence of chimera states, but also significantly enlarges their basin of attraction. 
When such a barrier is introduced into a ring network, which is in the synchronous state, a chimera state can emerge from the synchronous oscillations. 
The mechanism involves small phase differences that emerge at the border of the barrier, drift outwards and accumulate. 
After the threshold parameter within the barrier had been reset to its original value in the oscillatory regime, the generated chimera remained. 
This way of generating a chimera state is different from common procedures in the sense that there is no need to utilize random or specially prepared initial conditions.

The effects described in this work provide an interesting outlook on future lines of research and even applications. 
A better understanding of the effects that different `species' of nodes can have on chimera states offers promising options to modify local system parameters in order to achieve desired states. 
One possible example could be a targeted medication treatment to achieve incoherent, that is, asynchronous, dynamics of nerve cells. 
The coupling topology of the network, which would be hard, undesirable or even impossible to change, can be left untouched.

\begin{acknowledgments}
 This work was supported by the German Academic Exchange Service DAAD and the Greek State Scholarship Foundation IKY within the PPP-IKYDA framework. 
 PH acknowledges support by BMBF (grant no.  01Q1001B) in the framework of BCCN Berlin (Project A13). 
 TI and PH acknowledge support by Deutsche Forschungsgemeinschaft in the framework of SFB 910.
 JH acknowledges support by the EU/FP7-REGPOT-2012-2013-1 under grant agreement n316165. 
\end{acknowledgments}

\appendix

\section{Protocol for measuring the chimera drift velocity}\label{sec:prot-meas-chim}
Here we describe the procedure of measuring the change of chimera size and position.
All this applies for a fixed pair of parameters $(R,\sigma)$ at which a (1-) chimera exists.

The following are parameters to be supplied to the algorithm: 
$t_\text{avg}$ is the time that the moving average over the local order parameter is performed, 
$t_\text{check}$ is the time interval between the checkpoints at which the location of the chimera is calculated.
$n$  is the number of checkpoints used in the calculation of the speed and growth rate.
$Z_\text{thresh}$ is used for locating the incoherent region.
$\dot c_\text{thresh}$ and $\dot w_\text{thresh}$ are the threshold values for speed and growth rate respectively.
We used 
\begin{align*}
t_\text{avg}    &= 400,   & t_\text{check}        &= 200,           &   n                  &= 15,  \\
Z_\text{thresh} &= 0.02,  & \dot c_\text{thresh}  &= 5\cdot10^{-4},  & \dot w_\text{thresh} &= 1\cdot10^{-3}
\end{align*}
for the generation of Fig.~\ref{fig:control_efficiency}.
\\
\\{\bf Algorithm steps:}
\begin{compactenum}[1.]
\item[-1.] Generation of initial conditions.
   \begin{compactenum}[(i)]
   \item Generate a chimera state,
   \item locate the incoherent region,
   \item perform an index shift such that the incoherent region is placed exactly opposite from barrier.
   \end{compactenum}
\item[0.] Run the system with a barrier using those initial conditions from $t_0=0$ to $t_1=t_\text{avg}$. Set $j=1$
\item     Calculate $\langle Z_i \rangle_{\Delta t=t_\text{avg}} (t_j)$ as in Eq.~\eqref{eq:moving_average_of_Z}.
\item     Calculate the incoherent regions' borders $i_\text{left},\,i_\text{right}$ by using only $\langle Z_i \rangle_{\Delta t=t_\text{avg}} (t_j)$. 
           Borders are given by those indices $i$, where a threshold value of $1-Z_\text{thresh}$ is undercut.
\item     Calculate the barycentric coordinates $c_j,\,w_j$ of the incoherent region.
          $c_j = \left(i_\text{left} + i_\text{right}\right)/2$ and $w_j = i_\text{left} - i_\text{right}$. (Every operation to be understood modulo $N$).
\item     If $j\ge n$:
  \begin{compactenum}[(i)]
    \item Perform a linear regression over $(c_k,t_k)$ and $(w_k,t_k)$ with $k\in[j-n+1, j]$. Results are: speed $\dot c$ and growth rate $\dot w$.
    \item If both $\dot c$ and $\dot w$ undercut the threshold values $\dot c_\text{thresh}$ and $\dot w_\text{thresh}$: \\
           Go to 7.
  \end{compactenum}
\item     Run from $t_j$ to $t_{j+1} = t_j + t_\text{check}$. Set $j=j+1$
\item     Start over from 1.
\item     Print results
\begin{align*}
\dot c_\text{final} &= \dot c                \\ c_\text{final} &= \tilde c_0 + t_j \cdot \dot c_\text{final} \\
w_\text{final}      &=  w_\text{final} = \tilde w_0 + t_j \cdot \dot w_\text{final}, 
\end{align*}
where $\tilde c_0$ and $\tilde w_0$ are the intercepts obtained from the linear regression. 
\end{compactenum}

\section{Algorithm for detecting chimera states and the borders of incoherent regions}\label{sec:algor-detect-chim}
To the best of our knowledge, no detailed description on how to automatically and reliably detect and exactly locate the (multiple) incoherent regions in a (multi-) chimera state exists.
This is the reason why we present the algorithm we have developed in detail.

The main characteristic of a chimera state is the simultaneous presence of coherent and incoherent regions. 
The two main measures to tell these two regions apart are 
\begin{inparaenum}[(a)]
\item the average local order parameter $\langle Z_i\rangle_{\Delta t}$, see Eq.~\eqref{eq:local_order_parameter} and
\item the mean phase velocity $\omega_i$, see Eq.~\eqref{eq:average_phase_velocity}.
\end{inparaenum}

It is possible to use each of these measures alone to tell coherent and incoherent regions apart.  
There are however some disadvantages in doing so.
When using only the average local order parameter $\langle Z_i \rangle_{\Delta t}$, the disadvantages are that
\begin{inparaenum}[(i)]
\item the localization is not precise due to the spatial window used to calculate $Z_i$ and
\item regions displaying  wave-like behavior (which should be considered coherent) exist but can have a significantly lowered local order parameter due to the phase shift.
\end{inparaenum}

When using only the mean phase velocity $\omega_i$, the disadvantages are that
\begin{inparaenum}[(i)]
\item a smooth resolution of $\omega_i$ needs a very large time for averaging and
\item there can be (considering reasonable finite averaging times) quasi coherent regions with some variations in $\omega_i$, which would be labeled incoherent.
\end{inparaenum}

The advantages of using the averaged local order parameter is that it is sufficient to calculate it with a short time averaging window.
(Some temporal averaging is needed especially in the case of relaxation oscillators where the phase velocity can depend strongly on the phase.)
The advantage of using the mean phase velocity is its good spatial resolution.
To sum it up, the difference in using these two measures boils down to spatial vs. temporal resolution.

In addition, we use the mean phase velocity beforehand to check whether there is a chimera at all.
In a chimera state, the mean phase velocity is constant in coherent regions and varies in the incoherent region. 
In the synchronous state, the mean phase velocity is the same everywhere.
So, if the difference between minimum and maximum of the mean phase velocity is below a certain threshold, the algorithm is stopped immediately, as there is (most likely) no chimera state.

We propose to combine the advantages of the two measures in the following algorithm:
\begin{compactenum}
\item Calculate $\omega_i$ and $\langle Z_i \rangle_{\Delta t}$ from the timeseries. To simplify the notation, we write $Z_i$ instead of $\langle Z_i \rangle_{\Delta t}$ from now on.
\item If $\max_i (\omega_i) - \min_i(\omega_i) < \omega_\text{ex}$, stop.
\item Calculate $\omega_\text{coh} :=  \langle \omega_i \rangle_{\{i:Z_i \ge 1-Z_\text{thresh}\}\cap\{i:a_i=a_\text{osc}\} }$
\item Smooth $\omega_i$ by calculating \\ $\bar\omega_i = \left( \omega_{i-1} + \omega_i + \omega_{i+1}\right)/3$ (indices modulo $N$).
\item From here, only apply to oscillatory units. By using the two conditions 
\begin{align*} 
\mathcal{C}_i^\omega & := \bar\omega_i \le \omega_\text{coh} + \omega_\text{thresh} \\
\mathcal{C}_i^Z    &:= Z_i \ge 1-Z_\text{thresh},
\end{align*}
define the sets
\begin{align*}
\mathcal{S}_{+}  &:=\left\{ i : \mathcal{C}_i^\omega \wedge \mathcal{C}_i^Z\right\}                                 \\
\mathcal{S}_{-}  &:=\left\{ i : \neg \mathcal{C}_i^\omega \wedge \neg \mathcal{C}_i^Z\right\}                       \\
\mathcal{S}_{+-} &:=\left\{ i : \neg \mathcal{C}_i^\omega \vee   \neg \mathcal{C}_i^Z\right\} \cap \mathcal{S}_{-}
\end{align*}
Note that the sets are mutually exclusive by definition and that the union of them comprises all $i$.
\item Every unit $i$ that is in $\mathcal{S}_{+}$ is considered to belong to a coherent region, every unit $i$ that is in $\mathcal{S}_{-}$ is considered to belong to an incoherent region.
\item Process every connected component of $\mathcal{S}_{+-}$ according to the following scheme taking into account the neighboring regions and putting every $i$ in the pertaining component of $\mathcal{S}_{+-}$ in either $\mathcal{S}_+$ or $\mathcal{S}_-$:\\
\begin{tabular}{|c|c|c|c|c|c|c|} 
\cline{1-3}\cline{5-7}
 $\mathcal{S}_+$   & $\mathcal{S}_{+-}$ & $\mathcal{S}_+$  &   $\quad\rightarrow\quad$  &  $\mathcal{S}_+$ & $\mathcal{S}_+$   & $\mathcal{S}_+$  \\
\cline{1-3}\cline{5-7}
 $\mathcal{S}_-$   & $\mathcal{S}_{+-}$ & $\mathcal{S}_+$  &   $\quad\rightarrow\quad$  &  $\mathcal{S}_-$ & $\mathcal{S}_-$   & $\mathcal{S}_+$   \\
\cline{1-3}\cline{5-7}
 $\mathcal{S}_+$   & $\mathcal{S}_{+-}$ & $\mathcal{S}_-$  &   $\quad\rightarrow\quad$  &  $\mathcal{S}_+$ & $\mathcal{S}_-$   & $\mathcal{S}_-$   \\
\cline{1-3}\cline{5-7}
 $\mathcal{S}_-$   & $\mathcal{S}_{+-}$ & $\mathcal{S}_-$  &   $\quad\rightarrow\quad$  &  $\mathcal{S}_-$ & $\mathcal{S}_-$   & $\mathcal{S}_-$   \\
\cline{1-3}\cline{5-7}
\end{tabular}
\item Voilà, every $i$ that identifies an oscillatory unit is marked to belong either to a coherent ($\mathcal{S}_+$) or to an incoherent region ($\mathcal{S}_-$).
\end{compactenum}

Several parameters that need some adjustment were introduced in the various steps:
\begin{compactenum}
\item[In 1.] The spatial window $\delta$ used to calculate the local order parameter $Z_i$.
\item[In 1.] The time window $\Delta t$ over which $\omega_i$ and $\langle Z_i \rangle_{\Delta t}$ are calculated.
      In principle, one could assign two different timewindows for this task but there is no reason for doing so.
\item[In 2.] The existence threshold value $\omega_\text{ex}$.
\item[In 3.] The threshold value $Z_\text{thresh}$ used for calculating the frequency of the coherent region(s) (and in the next step for the second incoherence condition).
\item[In 5.] The threshold value $\omega_\text{thresh}$ for the first incoherence condition.
\end{compactenum}
The values we used for generating Fig.~\ref{fig:num_chim} are:
\begin{align*}
\delta          &= 25,    & \Delta t             &= 5000, &
Z_\text{thresh} &= 0.04,    & \omega_\text{thresh} &= 0.02 \\
\omega_\text{ex}&=0.05
\end{align*}
In principle, one could use also two different values for  $Z_\text{thresh}$ in steps 1 and 4, however practice showed that this brings no significant advantage. 

The advantage of using $Z_i$ as well as $\omega_i$ arises in step 6. 
By setting  $Z_\text{thresh}$ to a more restrictive (i.e. higher) value, the frequency of the coherent regions will be calculated more precisely. 
In step 6, the deviation from this precise value is used to identify the border parts of an incoherent region by the deviation from $\omega_\text{coh}$ when the local order parameter is already above the coherence threshold.

\bibliographystyle{prwithtitle}
\bibliography{ref}

\begin{thebibliography}{10}
\expandafter\ifx\csname url\endcsname\relax
  \def\url#1{{\tt #1}}\fi
\expandafter\ifx\csname urlprefix\endcsname\relax\def\urlprefix{URL }\fi

\bibitem{KUR02a}
Y.~Kuramoto and D.~Battogtokh: {\em {Coexistence of Coherence and Incoherence
  in Nonlocally Coupled Phase Oscillators.}\/}, Nonlin. Phen. in Complex Sys.
  {\bf 5}, 380 (2002).

\bibitem{ABR04}
D.~M. Abrams and S.~H. Strogatz: {\em Chimera states for coupled
  oscillators\/}, Phys.~Rev.~Lett. {\bf 93}, 174102 (2004).

\bibitem{OLM11}
S.~Olmi, A.~Politi, and A.~Torcini: {\em Chimera states and collective chaos in
  pulse-coupled neural networks\/}, BMC Neuroscience {\bf 12}, P336 (2011).

\bibitem{OME13}
I.~Omelchenko, O.~E. Omel'chenko, P.~H\"ovel, and E.~Sch{\"o}ll: {\em When
  nonlocal coupling between oscillators becomes stronger: patched synchrony or
  multichimera states\/}, Phys. Rev. Lett. {\bf 110}, 224101 (2013).

\bibitem{HIZ13}
J.~Hizanidis, V.~Kanas, A.~Bezerianos, and T.~Bountis: {\em Chimera states in
  networks of nonlocally coupled hindmarsh-rose neuron models\/}, Int. J.
  Bifurcation Chaos {\bf 24}, 1450030 (2014).

\bibitem{VUE14a}
A.~V{\"u}llings, J.~Hizanidis, I.~Omelchenko, and P.~H\"ovel: {\em Clustered
  chimera states in systems of type-{I} excitability\/}, New J.~Phys. {\bf 16},
  123039 (2014).

\bibitem{OME15}
I.~Omelchenko, A.~Provata, J.~Hizanidis, E.~Sch{\"o}ll, and P.~H\"ovel: {\em
  Robustness of chimera states for coupled {FitzHugh-Nagumo} oscillators\/},
  Phys. Rev. E {\bf 91}, 022917 (2015).

\bibitem{HIZ15}
J.~Hizanidis, E.~Panagakou, I.~Omelchenko, E.~Sch{\"o}ll, P.~H\"ovel, and
  A.~Provata: {\em Chimera states in population dynamics: networks with
  fragmented and hierarchical connectivities\/}, Phys. Rev. E {\bf 92}, 012915
  (2015).

\bibitem{OME11}
I.~Omelchenko, Y.~Maistrenko, P.~H\"ovel, and E.~Sch{\"o}ll: {\em Loss of
  coherence in dynamical networks: spatial chaos and chimera states\/}, Phys.
  Rev. Lett. {\bf 106}, 234102 (2011).

\bibitem{OME12}
I.~Omelchenko, B.~Riemenschneider, P.~H\"ovel, Y.~Maistrenko, and
  E.~Sch{\"o}ll: {\em Transition from spatial coherence to incoherence in
  coupled chaotic systems\/}, Phys. Rev.~E {\bf 85}, 026212 (2012).

\bibitem{OME15a}
I.~Omelchenko, A.~Zakharova, P.~H\"ovel, J.~Siebert, and E.~Sch{\"o}ll: {\em
  Nonlinearity of local dynamics promotes multi-chimeras\/}, Chaos {\bf 25},
  083104 (2015).

\bibitem{SCH14a}
L.~Schmidt, K.~Sch{\"o}nleber, K.~Krischer, and V.~Garcia-Morales: {\em
  Coexistence of synchrony and incoherence in oscillatory media under nonlinear
  global coupling\/}, Chaos {\bf 24}, 013102 (2014).

\bibitem{ZAK14}
A.~Zakharova, M.~Kapeller, and E.~Sch{\"o}ll: {\em Chimera death: Symmetry
  breaking in dynamical networks\/}, Phys.~Rev.~Lett. {\bf 112}, 154101 (2014).

\bibitem{SCH15a}
L.~Schmidt and K.~Krischer: {\em Clustering as a prerequisite for chimera
  states in globally coupled systems\/}, Phys. Rev. Lett. {\bf 114}, 034101
  (2015).

\bibitem{HAU15}
S.~W. Haugland, L.~Schmidt, and K.~Krischer: {\em Self-organized alternating
  chimera states in oscillatory media\/}, Scientific Reports {\bf 5}, 9883
  (2015).

\bibitem{OLM15}
S.~Olmi, E.~A. Martens, S.~Thutupalli, and A.~Torcini: {\em Intermittent
  chaotic chimeras for coupled rotators\/}, Phys. Rev. E {\bf 92}, 030901(R)
  (2015).

\bibitem{OME10a}
O.~E. Omel'chenko, M.~Wolfrum, and Y.~Maistrenko: {\em Chimera states as
  chaotic spatiotemporal patterns\/}, Phys. Rev.~E {\bf 81}, 065201(R) (2010).

\bibitem{OME12a}
O.~E. Omel'chenko, M.~Wolfrum, S.~Yanchuk, Y.~Maistrenko, and O.~Sudakov: {\em
  Stationary patterns of coherence and incoherence in two-dimensional arrays of
  non-locally-coupled phase oscillators\/}, Phys. Rev.~E {\bf 85}, 036210
  (2012).

\bibitem{PAN13}
M.~J. Panaggio and D.~M. Abrams: {\em Chimera states on a flat torus\/}, Phys.
  Rev. Lett. {\bf 110}, 094102 (2013).

\bibitem{PAN15a}
M.~J. Panaggio and D.~M. Abrams: {\em Chimera states on the surface of a
  sphere\/}, Phys. Rev. E {\bf 91}, 022909 (2015).

\bibitem{WOL11a}
M.~Wolfrum, O.~E. Omel'chenko, S.~Yanchuk, and Y.~Maistrenko: {\em Spectral
  properties of chimera states\/}, Chaos {\bf 21}, 013112 (2011).

\bibitem{OME13a}
O.~E. Omel'chenko: {\em Coherence-incoherence patterns in a ring of non-locally
  coupled phase oscillators\/}, Nonlinearity {\bf 26}, 2469 (2013).

\bibitem{HAG12}
A.~M. Hagerstrom, T.~E. Murphy, R.~Roy, P.~H\"ovel, I.~Omelchenko, and
  E.~Sch{\"o}ll: {\em Experimental observation of chimeras in coupled-map
  lattices\/}, Nature Physics {\bf 8}, 658 (2012).

\bibitem{TIN12}
M.~R. Tinsley, S.~Nkomo, and K.~Showalter: {\em Chimera and phase cluster
  states in populations of coupled chemical oscillators\/}, Nature Physics {\bf
  8}, 662 (2012).

\bibitem{MAR13}
E.~A. Martens, S.~Thutupalli, A.~Fourri{\`e}re, and O.~Hallatschek: {\em
  Chimera states in mechanical oscillator networks\/}, Proc. Nat. Acad.
  Sciences {\bf 110}, 10563 (2013).

\bibitem{WIC13}
M.~Wickramasinghe and I.~Z. Kiss: {\em Spatially organized dynamical states in
  chemical oscillator networks: Synchronization, dynamical differentiation, and
  chimera patterns\/}, PLoS ONE {\bf 8}, e80586 (2013).

\bibitem{LAR13}
L.~Larger, B.~Penkovsky, and Y.~Maistrenko: {\em Virtual chimera states for
  delayed-feedback systems\/}, Phys. Rev. Lett. {\bf 111}, 054103 (2013).

\bibitem{LAZ15}
N.~Lazarides, G.~Neofotistos, and G.~Tsironis: {\em Chimeras in squid
  metamaterials\/}, Phys.~Rev.~B {\bf 91}, 054303 (2015).

\bibitem{SIE14c}
J.~Sieber, O.~E. Omel'chenko, and M.~Wolfrum: {\em Controlling unstable chaos:
  Stabilizing chimera states by feedback\/}, Phys. Rev. Lett. {\bf 112}, 054102
  (2014).

\bibitem{BIC15}
C.~Bick and E.~A. Martens: {\em Controlling chimeras\/}, New J.~Phys. {\bf 17},
  033030 (2015).

\bibitem{FIT61}
R.~FitzHugh: {\em Impulses and physiological states in theoretical models of
  nerve membrane\/}, Biophys. J. {\bf 1}, 445 (1961).

\bibitem{NAG62}
J.~Nagumo, S.~Arimoto, and S.~Yoshizawa.: {\em An active pulse transmission
  line simulating nerve axon.\/}, Proc. IRE {\bf 50}, 2061 (1962).

\end{thebibliography}

\end{document}